\documentclass[12pt]{article}
\usepackage{amsmath,amssymb}
\usepackage{graphicx}
\usepackage[hypertex]{hyperref}

\numberwithin{equation}{section}

\def\cN{{\mathcal N}}
\def\cO{{\mathcal O}}

\def\ssp{\hspace{0.3mm}}
\def\({\left(}
\def\){\right)}
\renewcommand{\[}{\left[}
\renewcommand{\]}{\right]}

\def\eK{\mathrm{\bf K}}

\def\iom{i\hspace{0.2mm}\omega}

\def\bb#1{\mathbb{#1}}
\def\bmt#1{{\mbox{\boldmath$ #1 $}}}
\def\pare#1{\left\{ #1\right\}}

\def\abs#1{\left| #1\right|}

\def\RS#1{$\mathbb{R}_{\rm t} \! \times {\rm S}^{#1}$}

\def\AdSxS{{AdS${}_5 \times {}$S${}^5$}}
\def\RRP#1{$\mathbb{R}_{\rm t} \! \times \mathbb{RP}^{#1}$}
\def\RCP#1{$\mathbb{R}_{\rm t} \! \times \mathbb{CP}^{#1}$}
\def\AdSxP{{AdS${}_4 \times \mathbb{CP}^3$}}
\renewcommand{\eqref}[1]{$\({\rm \ref{#1}}\)$}

\begin{document}

{\ }
\vspace{-10mm}

\begin{flushright}
{\bf February 2009}\\[1mm]
{\small TCDMATH 09\,-\,07}\\[1mm]
\end{flushright}

\vskip 2cm

\begin{center}
\LARGE

\mbox{\bf Giant Magnons on $\bb{CP}^3$ by Dressing Method}


\vskip 2cm
\renewcommand{\thefootnote}{$\alph{footnote}$}

\large
\centerline{\sc Ryo Suzuki\footnotetext{{\tt rsuzuki@maths.tcd.ie}}}

\vskip 1cm

\emph{School of Mathematics, Trinity College, Dublin 2, Ireland}

\end{center}

\vskip 14mm

\centerline{\bf Abstract}

\vskip 6mm

We consider classical string spectrum of \RCP{3}, and construct a family of solutions with residual $SU(2)$ symmetry by the dressing method on $SU(4)/U(3)$ sigma model. All of them obey the square-root type dispersion relation often found in the theory with $su(2|2)$ symmetry. A single dyonic giant magnon is not found in this approach.

\vspace*{1.0cm}

\vfill
\thispagestyle{empty}
\setcounter{page}{0}
\setcounter{footnote}{0}
\renewcommand{\thefootnote}{\arabic{footnote}}
\newpage

\section{Introduction}\label{sec:intro}

An effective field theory of coincident membranes with $\cN=8$ superconformal symmetry in 1+2 dimensions is proposed by Bagger, Lambert, and Gustavsson (BLG) based on three-algebra \cite{Gustavsson07, BL07a, BL07b}. 
Aharony, Bergman, Jafferis and Maldacena (ABJM) proposed an $\cN=6$ superconformal Chern-Simons-matter theory with a tunable coupling constant $\lambda = N/k$ by generalizing the BLG theory to incorporate $U(N)_k \times U(N)_{-k}$ symmetry group --- which coincides with the special case of \cite{GW08,HLL08} --- and argued that their model at the 't Hooft limit is dual to type IIA superstring theory on the \AdSxP\ background \cite{ABJM08, BKKS08}.

The IIA on \AdSxP\ is less supersymmetric than the IIB on \AdSxS, which was conjectured to be dual to $\cN=4$ super Yang-Mills in 1+3 dimensions \cite{Maldacena97}. Integrability has provided us a powerful tool to study the AdS$_5$/CFT$_4$ correspondence, and a matter of central concern is whether and how similar techniques are applied to the AdS$_4$/CFT$_3$ case.

Despite huge and rapid progress on this subject, no conclusive answer has been given. Looking on the positive side, one finds the integrability of two-loop Hamiltonian in ABJM model \cite{MZ08, BR08, MSZ09, Zwiebel09}, classical integrability of superstring action (except for a subtle issue concerning strings in ${\rm AdS}_4$) \cite{AF08b, Stefanski08, GSW08},\footnote{The pure spinor action for \AdSxP\ is constructed in \cite{FG08, BGS08, DFGT08b}, which remains superconformal at quantum level as in the \AdSxS\ case \cite{Berkovits04f}.} and the proposal of all-loop Bethe Ansatz \cite{GV08b, AN08a, AN08b, AN09}, which is consistent with near-plane wave limit of string theory \cite{NT08, GGY08, GHO08, AGGHO08, Sundin08}. On the negative side, one finds disagreement between the one-loop energy of folded or circular string, and the proposed Bethe Ansatz \cite{MR08, AAB08, Krishnan08, GM08, MRT08}.

More data, especially the examples that are not found in \AdSxS\ case, are necessary to refine our understanding of the AdS$_4$/CFT$_3$ duality and its integrability \cite{GHOS08, LPP08, CW08, ABR08, BF08, LO08, AB08c, ABM08, BH09}. A good starting point will be to reconsider the correspondence between magnons in the asymptotic spin chain and giant magnon solutions on the decompactified worldsheet \cite{Beisert05, Beisert06b, HM06, GGY08}, as one could expect nice examples of the duality owing to the $su(2|2)$ symmetry.

The description by algebraic curve tells the classical string spectrum and its dispersion in a simple way \cite{GV08a, Shenderovich08}. Yet, to obtain further information such as (semiclassical) quantization and scattering \cite{CDO06b, Roiban06, Minahan07, PS07, CDM07},\footnote{There are proposals to perform quantization in the language of algebraic curve \cite{GV07b, Vicedo08a}.} it is useful to construct an explicit profile of the corresponding string solution.

The aim of this paper is to construct the explicit profile of classical strings on which only the existence and the dispersion have been known so far by means of algebraic curve. The relationship between a string solution and an algebraic curve is not explicit in general, so we have to construct the classical string solution from scratch. This sort of problem is quite difficult in general, due to the nonlinearity of differential equations. Here the integrability helps. In particular, the dressing method enables us to construct soliton solutions of integrable classical sigma models by means of linear algebra. The dressing method was developed in \cite{ZM78, ZS79, ZM80, HSS84a, HSS84b}. Application to Grassmannian sigma models including $\bb{CP}^N$ cases was intensively studied in \cite{Sasaki85, AP87, Piette88}. This method has also been applied to \RS{5}\ sigma model successfully in \cite{SV06, KSV06} to construct an explicit profile of multi giant magnon solutions.

Unfortunately, it is not guaranteed that the dressing method exhausts all solutions. In the dressing method, one chooses the vacuum solution and a particular embedding of $\bb{CP}^N$ into $SU(N)$ principal chiral model for some $N$. As is found by \cite{SV06, KSV06}, there are some solutions which can be obtained in one embedding, but cannot in another embedding.

In our case, we can obtain solutions with residual $SU(2)$ symmetry which carries only two nonvanishing components of angular momenta. However, we do not find a single ``small" dyonic solution \cite{Shenderovich08}, the one living on the $\bb{CP}^2$ subspace, and smoothly connected to $\bb{CP}^1$ giant magnons. We need to refine the dressing method to construct $\bb{CP}^2$ dyonic giant magnons.

The paper is organized as follows. We explain some preliminaries in Section \ref{sec:sigma}. In Section \ref{sec:dressing} we construct the solutions by dressing $SU(4)/U(3)$ model and discuss the limitations of our approach. Section \ref{sec:summary} is for summary and discussions. In Appendix \ref{sec:CPN Pohlmeyer}, we discuss how to relate the string motion with some sine-Gordon fields by means of Pohlmeyer-Lund-Regge reduction. We briefly comment on the dressing of $SO(6)/U(3)$ model in Appendix \ref{sec:so(6)/u(3)}.

\bigskip\noindent
{\bf Note added}: While the paper is in preparation, we find the paper \cite{HM09, KSV09} on arXiv, which has an overlap with our result. Their results seem to be equivalent to \eqref{gen dispersion}.

\section{Sigma model on $\bb{CP}^3$}\label{sec:sigma}

\subsection{Action in conformal gauge}

We consider a classical $\sigma$-model on \RCP{3}, particularly the one obtained by large $k$ limit of ${\rm S}^7/\bb{Z}_k$\,. When the $\bb{Z}_k$ action becomes gauging of ${\rm S}^1$, we obtain the $\bb{CP}^3$ space. We set the radius of ${\rm S}^7$ to unity, and introduce an embedding ${\rm S}^7 \subset \bb{C}^4$. The coordinates on $\bb{C}^4$ can be identified as the homogeneous coordinates of the $\bb{CP}^3$, normalized as
\begin{equation}
\sum_{i=1}^{4} \bar z_i \, z_i = 1.
\label{CP3 norm}
\end{equation}
Since the $\bb{CP}^3$ space has local scale invariance, we must identify two points $z_i$ and $c \ssp z_i$ with $c \in \bb{C}$. The condition \eqref{CP3 norm} partially fixes the gauge, and the residual symmetry is $U(1)$.

The $\sigma$-model action in conformal gauge compatible with \eqref{CP3 norm} is given by
\begin{equation}
S = - 2 h \int d^2 \sigma \[ \gamma^{\alpha \beta} \Bigl\{ - \frac14 \(\partial_\alpha t\) \(\partial_\beta \ssp t\) + ( D_\alpha z_i )^\dagger \, D_\beta \ssp z_i \Bigr\} + \Lambda \( \abs{z_i}^2 - 1\) \],
\label{CP3 action}
\end{equation}
where $h$ is the string tension,\footnote{$h (\lambda) = \sqrt{\lambda/2} + \cdots$ for $\lambda \gg 1$.} $\Lambda$ a Lagrange multiplier. The worldsheet metric is $\gamma^{\tau \tau} = - \gamma^{\sigma \sigma} = -1,\ \gamma^{\sigma \tau} = \gamma^{\tau \sigma} = 0$. Covariant derivatives acts on fields as
\begin{equation}
D_\alpha X = \partial_\alpha X - i A_\alpha X, \qquad \overline{D_\alpha} X = \partial_\alpha X + i A_\alpha X.
\label{def:deriv}
\end{equation}
Path integral of $A_\alpha$ and $\Lambda$ gives,
\begin{equation}
A_\alpha = \frac{i \( z_i \, \partial_\alpha \bar z_i - \bar z_i \, \partial_\alpha z_i \)}{2 \bar z_i z_i} = - i \bar z_i \, \partial_\alpha z_i \,.
\label{U1 connection}
\end{equation}
The equations of motion are\footnote{If one expands the covariant the derivative, one finds \begin{equation} D_\alpha^2 z_i = \partial_\alpha^2 z_i - 2 i A^\alpha \partial_\alpha z_i - i (\partial_\alpha A^\alpha) z_i - A_\alpha^2 z_i \,,\qquad \abs{D_\alpha z_j}^2 = \abs{\partial_\alpha z_j}^2 - A_\alpha^2 \,. \end{equation}}
\begin{equation}
0 = D_\alpha^2 z_i + \abs{D_\alpha z_j}^2 z_i \,,
\label{CP3 eom}
\end{equation}
and Virasoro constraints,
\begin{alignat}{2}
\frac{\kappa^2}{4} &= \abs{D_\sigma z_i}^2 + \abs{D_\tau z_i}^2 & &= \abs{\partial_\sigma z_i}^2 + \abs{\partial_\tau z_i}^2 - \( A_\sigma^2 + A_\tau^2 \),
\label{CP3 virasoro 1} \\
0 &= {\rm Re} \pare{ ( D_\sigma z_i )^\dagger \, D_\tau z_i } & &= {\rm Re} \pare{ \partial_\sigma \bar z_i \, \partial_\tau z_i } - A_\sigma A_\tau \,,
\label{CP3 virasoro 2}
\end{alignat}
for $t = \kappa \tau$. Any solution of the equations of motion and Virasoro constraints on $\bb{CP}^m \ (m < 3)$ subspace can be lifted to a solution of $\bb{CP}^3$ \cite{AA08}.

In general, the $U(1)$ field strength is not zero:
\begin{equation}
F_{\alpha \beta} \equiv \partial_\alpha A_\beta - \partial_\beta A_\alpha = - i \( \partial_\beta \bar z_i \, \partial_\alpha z_i - \partial_\beta z_i \, \partial_\alpha \bar z_i \) \neq 0,
\label{def:Fpm}
\end{equation}
and therefore one cannot achieve $A_\tau = A_\sigma = 0$ by any gauge transformation.

The conserved charges are defined by
\begin{equation}
E = h \int d \sigma \( \partial_\tau t \),\qquad
J_{\ell m} = 4 h \int d \sigma \; {\rm Im} \( \bar z_\ell \, D_\tau z_m \).
\label{CP3 charges}
\end{equation}
They satisfy $\sum_k J_{kk} = 0$. We use the notation $J_k = J_{kk}$ when the off-diagonal components of $J_{\ell m}$ are zero. This is indeed the case regarding all solutions discussed below.

\subsection{Coset embedding}

The $\bb{CP}^3$ sigma model can be embedded into $SU(N)$ principal chiral models in several ways. We have to choose one embedding to apply the dressing method. The simplest is $\bb{CP}^3 \subset SU(4)$. Let $\theta$ be an involution of $SL(4)$ given by $\theta = {\rm diag} \(1, -1, -1, -1 \)$, then $\bb{CP}^3$ is characterized by
\begin{equation}
\bb{CP}^3 = \frac{SU(4)}{S \[ U(3) \times U(1) \]} = \pare{ g \in SU(4) \, | \ \theta g \ssp \theta g = {\bf 1}_{4} },
\label{SU4 embedding}
\end{equation}
where ${\bf 1}_N$ is an $N \times N$ identity matrix. Equivalently, elements of $\bb{CP}^3$ can be written as
\begin{equation}
g \in \bb{CP}^3 \subset {\rm SU} (4) \quad \Longleftrightarrow \quad g = \theta \( {\bf 1}_{4} - 2 P \), \quad {\rm with} \quad P = P^2,\ P^\dagger = P.
\label{g projector}
\end{equation}

Suppose that the projector $P$ has rank one whose image is spanned by $z = \( z_1 , \ldots , z_{N+1} \)^{T}$. We identify this vector $z$ with the homogeneous coordinates of $\bb{CP}^N$ normalized to $z^\dagger z = 1$. The projector is written as
\begin{equation}
P_{ij} = z_i \, \bar z_j \,,
\label{projector z}
\end{equation}
which is manifestly gauge-invariant. From this explicit relation, one can rewrite the \RCP{N}\ action as \cite{EF80}\footnote{There is an identity $\[ (D_a z)^\dagger (D_b z) + (D_b z)^\dagger (D_a z) \] = {\rm tr} \[ \partial_a P \, \partial_b P \] = - \frac14 \, {\rm tr} \[ j_a \ssp j_b \]$.}
\begin{equation}
S = 2 h \int d^2 \sigma \pare{ - \frac14 \( \partial_a t \)^2 + \frac12 \, {\rm tr} \[ \( \partial_a P \)^2 \] + \Lambda \( P^2 - P \) }.
\end{equation}
The equation of motion is
\begin{equation}
\[ \partial_a^2 P, P \] = 0,
\label{eom projector}
\end{equation}
and Virasoro constraints are
\begin{equation}
{\rm tr} \[ \( \partial_\pm P \)^2 \] = \frac{\kappa^2}{2} \,, \qquad {\rm for} \quad t = \kappa \tau,
\label{Virasoro projector}
\end{equation}
where we introduced light-cone coordinates by $\tau = \sigma^+ + \sigma^-,\ \sigma = \sigma^+ - \sigma^-$.

We may rewrite the equation of motion as $\partial^a \[ \partial_a P, P \] = 0$; this is the conservation law for the currents
\begin{alignat}{7}
j_a &\equiv g^{-1} \partial_a g & &= - 2 \[ \partial_a P, P \] & &= 2 \[ z \(D_a z \)^\dagger - \( D_a z \) z^\dagger \],
\label{current l-inv} \\
\ell_a &\equiv - \partial_a g \, g^{-1} & &= - 2 \ssp \theta \[ \partial_a P, P \] \theta & &= 2 \theta \[ z \(D_a z \)^\dagger - \( D_a z \) z^\dagger \] \theta.
\label{current r-inv}
\end{alignat}
They define the conserved charges as,
\begin{equation}
J_{\ell m} \equiv i h \int d \sigma \( j_\tau \)_{\ell m} = - 2 i h \int d \sigma \[ \partial_\tau P, P \]_{\ell m} = 4 h \int d \sigma \; {\rm Im} \( \bar z_\ell \ssp D_\tau z_m \).
\end{equation}

\bigskip
The $\bb{CP}^3$ model can also be embedded into the $SU(6)$ principal chiral model, through $SO(6) \subset SU(6)$ and
\begin{equation}
\bb{CP}^3 \simeq \frac{SO(6)}{U(3)} = \pare{ g \in SO(6) \, | \ K g \ssp K g = - {\bf 1}_{6} },
\label{SO6 embedding}
\end{equation}
where $K$ is an antisymmetric involution of $SL(6)$. An explicit description of \eqref{SO6 embedding} is given in \cite{AF08b}. They found
\begin{equation}
g \equiv e^Y = {\bf 1} + \sin \rho \, \tilde Y + \(1 - \cos \rho\) \tilde Y^2,\qquad \tilde Y \equiv \frac{Y}{\rho}, \qquad \rho^2 = \sum_{i=1}^6 y_i^2 \,.
\end{equation}
where
\begin{equation}
Y = \begin{pmatrix}
0& 0 & y_1 & y_2 & y_3 & y_4 \\
0 & 0 & y_2 & -y_1 & y_4 & -y_3  \\
-y_1  & -y_2 & 0 & 0 & y_5 & y_6  \\
-y_2  & y_1 & 0& 0 & y_6 & -y_5 \\
-y_3  & -y_4 & -y_5 & -y_6 & 0 & 0 \\
-y_4  & y_3 & -y_6 & y_5 & 0& 0
\end{pmatrix}, \qquad K = \begin{pmatrix}
0 & 1 & 0 & 0 & 0 & 0 \\
-1 & 0 & 0 & 0 & 0 & 0  \\
0  & 0 & 0 & 1 & 0 & 0  \\
0  & 0 & -1& 0 & 0 & 0 \\
0  & 0 & 0 & 0 & 0 & 1 \\
0  & 0 & 0 & 0 & -1& 0
\end{pmatrix}.
\label{def:YK}
\end{equation}
The coordinates $\tilde y_i \equiv y_i/\rho$ are related to the normalized homogeneous coordinates on $\bb{CP}^3$ as follows:
\begin{equation}
\( z_1 , z_2 , z_3 , z_4 \) = \( \, \sin \rho \( \tilde y_1 + i \tilde y_2 \) , \sin \rho \( \tilde y_3 + i \tilde y_4 \) , \sin \rho \( \tilde y_5 + i \tilde y_6 \) , \cos \rho \, \).
\end{equation}

\bigskip
Finally, there is an embedding $\bb{RP}^3 \simeq SU(2) \times \overline{SU(2)} \subset SU(4)$, defined by
\begin{equation}
g = \begin{pmatrix}
g_2 & 0 \\
0 & \bar g_2
\end{pmatrix},\quad
g_2 = \sqrt 2 \begin{pmatrix}
z_1 & - i z_2 \\
- i \bar z_2 & \bar z_1
\end{pmatrix},\quad
\begin{pmatrix}
z_1 \\
z_2
\end{pmatrix} = \begin{pmatrix}
\bar z_4 \\
\bar z_3
\end{pmatrix},\quad
\sum_{j=1}^4 \abs{z_j}^2 = 1.
\end{equation}

\subsection{Examples of spectrum in the decompactification limit}

Soliton-like solutions on the $\bb{CP}^3$ sigma model can be found in the decompactification limit. In conformal gauge, this limit can be achieved by rescaling the worldsheet coordinates by $(\tilde \tau, \tilde \sigma) = (\mu \tau, \mu \sigma)$ with $\mu \to \infty$. Below we will discuss examples of the classical strings which obey the boundary condition
\begin{equation}
\( \frac{z_1}{z_4} \,, \frac{z_2}{z_4} \,, \frac{z_3}{z_4} \) \to \( e^{i t + i p_\pm} \,, 0, 0 \), \qquad p \equiv \frac{p_+ - p_-}{2} \,,
\label{CP3 boundary condition}
\end{equation}
as $\tilde \sigma \to \pm \infty$. See \cite{AA08} for a thorough discussion.

\subsubsection{Pointlike strings}

The simplest solution is the pointlike string, or the geodesic in $\bb{CP}^3$, given by
\begin{equation}
t = \omega \tau,\quad \( z_1 , z_2 , z_3, z_4 \) = \( \frac{1}{\sqrt 2} \, e^{i \omega \tau/2} \,, 0, 0, \frac{1}{\sqrt 2} \, e^{- i \omega \tau/2} \),
\label{pointlike string}
\end{equation}
whose conserved charges are
\begin{equation}
E = J_1 = - J_4 = h \, 2 \pi\omega.
\label{BPS relation string}
\end{equation}
Note that the profile $t = \omega \tau,\ \( z_1 , z_2 , z_3, z_4 \) = \( e^{i \omega \tau},0,0,0\)$ is a solution but meaningless, because the Virasoro constraints imposes $\omega=0$.

From the viewpoint of AdS/CFT, the pointlike string should correspond to the BPS states of the ABJM model,
\begin{equation}
\cO = {\rm tr} \, \[ Y^1 (Y_4)^\dagger Y^1 (Y_4)^\dagger \dots \].
\label{ABJM BPS}
\end{equation}
If $J_1,$ and $J_4$ are the numbers of $Y_1, Y_4$\,, the conformal dimension of \eqref{ABJM BPS} satisfies
\begin{equation}
\Delta = \frac{J_1 - J_4}{2} \,,
\label{BPS relation gauge}
\end{equation}
in agreement with \eqref{BPS relation string}. We can consider magnon excitations over the BPS vacuum. In the case of $SU(2) \times SU(2)$ sector, they are given by the replacement
\begin{equation}
Y^1 \ \to \ Y^2 \ {\rm or} \ Y^3, \qquad (Y_4)^\dagger \ \to \ (Y_3)^\dagger \ {\rm or} \ (Y_2)^\dagger.
\label{su2xsu2 magnons}
\end{equation}

\subsubsection{Recycling solutions of \bmt{\bb{R}_{\rm t} \! \times {\rm S}^3}}

Any classical string solution on ${\rm S}^3$ can be mapped to a solution on $\bb{RP}^3$. Let $(t_S \,, \xi_1 , \xi_2)$ be a classical string solution on \RS{3}\ with $\abs{\xi_1}^2 + \abs{\xi_2}^2 = 1$. Then the ansatz
\begin{equation}
t = 2 \ssp t_S \,, \qquad \( z_1 , z_2 , z_3, z_4 \) = \frac{1}{\sqrt 2} \( \xi_1, \xi_2 \,, \xi_2^* \,, \xi_1^* \),
\label{RP3 ansatz 1}
\end{equation}
is a consistent solution of \RRP{3}. The boundary condition \eqref{CP3 boundary condition} is translated into
\begin{equation}
\( \xi_1 \,, \xi_2 \) \to \( \exp \[ i t_S + \frac{i p_\pm}{2} \] , 0 \), \quad p \equiv \frac{p_+ - p_-}{2} \qquad {\rm as} \ \ \tilde \sigma \to \pm \infty.
\label{RS3 boundary condition}
\end{equation}
If $\xi_2$ is real, we can easily generalize the ansatz \eqref{RP3 ansatz 1} by using the $SU(2)$ symmetry acting on $(z_2 \,, z_3)$ \cite{ABR08, GHOS08}.

From strings living on \RS{2}, there are two ways to construct a consistent solution of \RCP{3}. One is to use the isomorphism $\bb{CP}^1 \simeq {\rm S}^2$, and the other is to use the ansatz \eqref{RP3 ansatz 1}. Roughly said, the energy of the $\bb{CP}^1$ (``small") solution is just a half of the $\bb{RP}^2$ (``big") solution.

\subsubsection{Giant magnons}

The profile of $\bb{RP}^3$ dyonic giant magnons in conformal gauge is given by
\begin{equation}
t = 2 \sqrt{1 + u_2^2} \; \tilde \tau \,,\quad 
z_1 = z_4^* = \frac{1}{\sqrt 2} \, \xi_1 \,,\quad 
z_2 = z_3^* = \frac{1}{\sqrt 2} \, \xi_2 \,,
\label{DGM on RP3}
\end{equation}
where \cite{CDO06a, OS06}
\begin{gather}
\xi_{1} = \frac{\sinh(X  - i \omega)}{\cosh(X )}
\, e^{i \tan(\omega)X + i u_1 T}\,,\qquad 
\xi_{2} = \frac{\cos (\omega)}{\cosh(X )} \;
\, e^{i u_2 T} \,,
\label{DGM on RS3} \\
T(\tau,\sigma) \equiv \frac{\tilde \tau - v \tilde \sigma}{\sqrt{1 - v^2}} \,,\quad
X(\tau,\sigma) \equiv \frac{\tilde \sigma - v \tilde \tau}{\sqrt{1 - v^2}} \,,\quad
v = \frac{\tan (\omega)}{u_1} \,,\quad
u_1^2 - u_2^2 = \frac{1}{\cos^2 \omega} \,.
\label{def:X,T}
\end{gather}
The boundary conditions \eqref{RS3 boundary condition} are satisfied if $p = \pi - 2 \omega$. The conserved charges are given by
\begin{alignat}{2}
&E & &= 4 h \, u_1 \( 1 - \frac{\tan^2 (\omega)}{u_1^2} \) \eK(1), \\
J_1 = - &J_4 & &= 4 h \, u_1 \[ \( 1 - \frac{\tan^2 (\omega)}{u_1^2} \) \eK(1) - \cos^2 (\omega) \], \\
J_2 = - &J_3 & &= 4 h \, u_2 \cos^2 (\omega),
\end{alignat}
where $\eK(1)$ is a divergent constant. They satisfy the relation
\begin{eqnarray}
E - \frac{J_1 - J_4}{2} = \sqrt{ \( \frac{J_2 - J_3}{2} \)^2 + 16 h^2 (\lambda) \sin^2 \( \frac{p}{2} \)} \,.
\end{eqnarray}

It is interesting to consider semiclassical quantization of the $\bb{RP}^3$ dyonic giant magnons following \cite{CDO06a}. If we observe the motion of the string \eqref{DGM on RP3} in the moving frame $\( \tilde z_1 , \tilde z_2, \tilde z_3, \tilde z_4 \) = \( z_1 \, e^{-i t/2}, z_2, z_3, z_4 \, e^{it/2} \)$, it becomes periodic with respect to $\tau$ when $\omega=0$, or $p=\pi$. This period measured by the AdS time $t$ is ${\cal T} = 4 \pi \coth \frac{q}{2}$\,, with $q$ defined by $\sinh \frac{q}{2} \equiv u_2 \cos (\omega)$. Then, the Bohr-Sommerfeld condition tells that the action variable
\begin{equation}
I = \frac{1}{2 \pi} \oint {\cal T} \, d \( E - \frac{J_1-J_4}{2} \)_{p=\pi} = J_2 - J_3 \,,
\label{Bohr-Sommerfeld condition}
\end{equation}
is an integer.

\bigskip
Next, we consider $\bb{CP}^1$ giant magnon. We define polar coordinates on $\bb{CP}^1$ by
\begin{equation}
\( z_1 \,, z_2 \,, z_3 \,, z_4 \) = \( \sin \frac{\theta}{2} \, e^{i \phi/2}, 0, 0, \cos \frac{\theta}{2} \, e^{-i \phi/2} \).
\label{def:CP1 polar}
\end{equation}
The $\bb{CP}^1$ giant magnon is given by
\begin{equation}
\cos \theta = \frac{\cos (\omega)}{\cosh (X)} \,, \quad e^{i \phi} = \sqrt{\frac{\sinh (X - \iom)}{\sinh (X + \iom)}} \; e^{i X \tan (\omega) + i T u} \,, \quad u = \frac{1}{\cos (\omega)} \,.
\label{polar GM}
\end{equation}
We can rewrite the solution in terms of $\bb{CP}^1$ homogeneous coordinates. Such expressions are not unique due to the $U(1)$ degree of freedom. If we look for the solution with $A_X=0$, we obtain
\begin{equation}
t = \frac{T + X \sin \omega}{\cos \omega} \,,\qquad z_1 = \frac{\sinh \(\frac{X-\iom}{2}\)}{\sqrt{\cosh (X)}} \, e^{it/2}, \qquad z_4 = \frac{\cosh\(\frac{X+\iom}{2}\)}{\sqrt{\cosh (X)}} \, e^{-it/2} \,,
\label{cp1gm tz}
\end{equation}
with $z_2 = z_3 = 0$. The $U(1)$ gauge fields defined in \eqref{U1 connection} are given by
\begin{equation}
A_X = 0,\qquad A_T = - \frac{1}{2 \cosh X} \,,
\end{equation}
which is in fact the Lorenz gauge $\partial^\alpha A_\alpha = 0$. The Lagrangian density is related to the kink solution of sine-Gordon model by Pohlmeyer-Lund-Regge reduction \cite{Pohlmeyer76, LR76}
\begin{equation}
- 4 \abs{D_\alpha z_i}^2 = 1 - \frac{2}{\cosh^2 (X)} \,.
\label{CP1 kink}
\end{equation}
We must set $v = \sin \omega$ in \eqref{cp1gm tz} to obtain $t = \tau$. The boundary condition \eqref{CP3 boundary condition} gives $p = \pi/2 - \omega$. The conserved charges satisfy the dispersion relation
\begin{equation}
E - \frac{J_1 - J_4}{2} = 2 h \abs{\cos \omega} = 2 h \abs{\sin p} \,.
\label{CP1 dispersion}
\end{equation}

\section{Dressing method}\label{sec:dressing}

We review the paper of Sasaki \cite{Sasaki85}, where he constructed solitons on the $\bb{CP}^N$ sigma model by dressing $SU(N+1)/U(N)$, and make a comment on the case of other embeddings.

\subsection{Dressing $\bmt{SU(N+1)/U(N)}$}

We begin with rewriting the equation of motion and the Bianchi identity in terms of a Lax pair, as
\begin{alignat}{4}
0 &= \pare{ \partial_+ - \frac{\partial_+ g \, g^{-1}}{1 + \lambda} } \psi & &= \pare{ \partial_+ - \frac{2 \ssp \theta \[ \partial_+ P, P \] \theta}{1+ \lambda} } \psi, \notag \\[2mm]
0 &= \pare{ \partial_- - \frac{\partial_- g \, g^{-1}}{1 - \lambda} } \psi & &= \pare{ \partial_- - \frac{2 \ssp \theta \[ \partial_- P, P \] \theta}{1- \lambda} } \psi.
\label{def:lax pair}
\end{alignat}
When $\lambda=0$, they are solved by
\begin{equation}
\psi (\lambda = 0) = g = \theta \( {\bf 1}_{N+1} - 2 P \).
\label{psi lambda0}
\end{equation}
The relation $j_a = \theta \ell_a \theta$ imposes an additional constraint on $\psi (\lambda)$. If we rewrite \eqref{def:lax pair} as\footnote{The derivative, {\it i.e.} the first term in the bracket, acts on anything that follows.}
\begin{equation*}
0 = g^{-1} \pare{ \partial_\pm - \frac{(\partial_\pm g) \, g^{-1}}{1 \pm \lambda} } g \( g^{-1} \psi \) = \pare{ \partial_\pm + \frac{g^{-1} (\partial_\pm g)}{1 \pm 1/\lambda} }  g^{-1} \psi = \theta \pare{ \partial_\pm - \frac{(\partial_\pm g) \, g^{-1}}{1 \pm 1/\lambda} } \theta g^{-1} \psi,
\end{equation*}
we find inversion symmetry
\begin{equation}
\psi (\lambda) = g \ssp \theta \ssp \psi (1/\lambda) \ssp \theta.
\label{inversion constraint}
\end{equation}
The unitarity condition on $\psi (\lambda)$ follows from \eqref{def:lax pair}, as
\begin{equation}
\[ \psi (\bar \lambda) \]^\dagger \psi (\lambda) = {\bf 1}_{N+1} \,.
\label{unitarity constraint}
\end{equation}
The right multiplication $\psi \to \psi \ssp U$ with a constant unitary matrix $U$ leaves the system of equations \eqref{def:lax pair} invariant. We can fix this ambiguity by $\psi (\lambda = \infty) = {\bf 1}_{N+1}$.

\bigskip
A classical solution of the $\bb{CP}^N$ model is a map from worldsheet to $\bb{CP}^N$ subject to certain constraints. Since its image lies within $SU(N+1)$, all solutions are related by some unitary transformations. We assume that solutions are meromorphic functions of $\lambda$, and try to extend the unitary transformation over the complex $\lambda$ plane. The dressing method provides us a simple way to construct such transformation matrices.

Let $\psi$ be the simplest solution of \eqref{def:lax pair}, \eqref{inversion constraint}, \eqref{unitarity constraint}, and $\tilde \psi = \chi \psi$ be another solution. We call $\chi$ the dressing matrix. If $\chi$ is not a constant matrix, then $\tilde \psi (\lambda=0) = \tilde g$ becomes a new solution of the $\bb{CP}^N$ model.

The dressing matrix for the $\bb{CP}^N$ sigma model has been constructed explicitly in \cite{Sasaki85} for the case of Euclidean worldsheet. The dressing matrix for Minkowski worldsheet can be obtained by replacing $( \lambda_1 \,, \bar \lambda_1 )$ with $( \lambda_1 \,, - \bar \lambda_1 )$ in the Euclidean result, because the light-cone coordinates are not complex conjugate with each other in the Minkowski case.

Let $\psi$ be the vacuum solution of \eqref{def:lax pair}. We introduce variables $g$ and $h$ by
\begin{equation}
\psi (\sigma^a, \lambda = 0) \equiv g = \theta \( {\bf 1} - 2 P \),\qquad \theta h = \psi (\sigma^a, \bar \lambda_1) \ssp u,
\end{equation}
where $u$ is a constant vector which parameterizes the dressed solution.\footnote{Is is possible to generalize $u$ into a constant, rectangular matrix following \cite{HSS84a, HSS84b}.\label{note:rank}} The dressing matrix of \cite{Sasaki85}, modified for our case is given by
\begin{alignat}{3}
\chi (\lambda) &= {\bf 1} + \frac{Q_1}{\lambda - \lambda_1} + \frac{Q_2}{\lambda - 1/\lambda_1} \,,
\label{general dressing mat} \\[1mm]
Q_1 &= \frac{\lambda_1}{\Lambda} \( \frac{\bar \lambda_1}{\lambda_1 - \bar \lambda_1} \, \theta h \beta h^\dagger \theta + \frac{1}{1 - \bar \lambda_1 \lambda_1} \, g h \gamma h^\dagger \theta \),
\label{dress Q1} \\[1mm]
Q_2 &= \frac{1}{\Lambda} \( - \frac{1}{\lambda_1 - \bar \lambda_1} \, g h \beta h^\dagger g^\dagger + \frac{\bar \lambda_1}{1 - \bar \lambda_1 \lambda_1} \, \theta h \gamma h^\dagger g^\dagger \),
\label{dress Q2}
\end{alignat}
where $\Lambda, \beta, \gamma$ are real numbers defined by
\begin{equation}
\Lambda = \bar \lambda_1 \lambda_1 \( \frac{\beta^2}{(\lambda_1 - \bar \lambda_1)^2} - \frac{\gamma^2}{(1 - \bar \lambda_1 \lambda_1)^2} \), \quad \beta = h^\dagger h, \quad \gamma = h^\dagger \theta g h.
\label{def:LBG}
\end{equation}
Since the dressed solution $\tilde g = \chi(0) g$ satisfies $( \theta \tilde g )^2 = {\bf 1}$ and $\tilde g \tilde g^\dagger = {\bf 1}$, we can introduce the dressed projector by $\tilde g = \theta ( {\bf 1} - 2 \tilde P )$. The dressed projector takes the form
\begin{multline}
\tilde P = P - \frac{1}{2 \Lambda} \Bigg\{ \frac{\bar \lambda_1 h \beta h^\dagger ({\bf 1} - 2 P) - \lambda_1 ({\bf 1} - 2 P) h \beta h^\dagger}{\lambda_1 - \bar \lambda_1}  \\
+ \frac{({\bf 1} - 2 P) h \gamma h^\dagger ({\bf 1} - 2 P) - \bar \lambda_1 \lambda_1 h \gamma h^\dagger}{1 - \bar \lambda_1 \lambda_1}  \Bigg\},
\end{multline}
and its image, $\tilde P \tilde z = \tilde z$, is given by
\begin{equation}
\tilde z = \frac{1}{\Lambda_z} \( \alpha z + h h^\dagger z \), \quad
\alpha = - \frac{\lambda_1 \beta}{\lambda_1 - \bar \lambda_1} + \frac{\gamma}{1 - \bar \lambda_1 \lambda_1} \,,\quad
\Lambda_z = \sqrt{ \abs{\alpha}^2 + \frac{2 \gamma}{1 - \bar \lambda_1 \lambda_1} \, \abs{h^\dagger z}^2 }.
\label{general dressed solution}
\end{equation}
This gives us the dressed solution in terms of the normalized homogeneous coordinates of $\bb{CP}^N$.

\subsection{Dressed solution on $\bb{CP}^3$}

Let us concentrate on the $\bb{CP}^3$ case. We choose vacuum as the following BPS solution:\footnote{Below we shall use $(\tau, \sigma)$ in place of $(\tilde \tau, \tilde \sigma) = (\mu \tau, \mu \sigma)$.}
\begin{equation}
t = \tau, \qquad z = \frac{1}{\sqrt 2} \begin{pmatrix}
e^{i \tau/2} \\
0 \\
0 \\
e^{-i \tau/2}
\end{pmatrix}, \qquad
g = \begin{pmatrix}
0 && 0 && 0 && - e^{i \tau} \\
0 && -1 && 0 && 0 \\
0 && 0 && -1 && 0 \\
e^{-i \tau} && 0 && 0 && 0
\end{pmatrix}.
\label{cp3 vac g}
\end{equation}
We parametrize the initial vectors $u$ and $h$ by
\begin{gather}
u = \rho_u \begin{pmatrix}
e^{i \nu_1} \, \cos \rho_1 \cos \rho_2 \\
e^{i \nu_2} \, \sin \rho_1 \cos \rho_3 \\
e^{i \nu_3} \, \sin \rho_1 \sin \rho_3 \\
e^{i \nu_4} \, \cos \rho_1 \sin \rho_2
\end{pmatrix}, \qquad
h = \rho_u \begin{pmatrix}
- e^{i \nu_4 + i \Sigma (\bar \lambda_1)} \, \cos \rho_1 \sin \rho_2 \\
e^{i \nu_2} \, \sin \rho_1 \cos \rho_3 \\
e^{i \nu_3} \, \sin \rho_1 \sin \rho_3 \\
- e^{i \nu_1 - i \Sigma (\bar \lambda_1) } \, \cos \rho_1 \cos \rho_2
\end{pmatrix}, 
\label{cp3 h1} \\[1mm]
\Sigma (\lambda, \sigma^+, \sigma^-) = \frac{\sigma^+}{1 + \lambda} + \frac{\sigma^-}{1 - \lambda} \,,
\end{gather}
The parametrization \eqref{cp3 h1} can be simplified. Multiplication by a complex constant $u \to c \ssp u$ does not modify the dressing matrix; we set $\rho_u = 1,\ \nu_1 + \nu_4 = 0$. Two real degrees of freedom of $u$ go away by constant shifts of $\sigma^\pm$. Let $\Sigma_0$ be the displacement of $\Sigma (\bar \lambda_1 , \sigma^+, \sigma^-)$ after such shifts. The translation $(\nu_1 \,, \nu_4) \to (\nu_1 - \delta \nu \,, \nu_4 + \delta \nu)$ can be cancelled by the real part of $\Sigma_0$\,, and a particular combination of shifts on $\rho_1 \,, \rho_2 \,, \rho_u$ is cancelled by the imaginary part; we may set $\nu \equiv \nu_4 - \nu_1 = 0$ and $\rho_2 = \pi/4$.

Under a global $SU(4)$ rotation, the vectors $z, h, \theta g h$ behave in the same manner:
\begin{equation}
z \ \to O \ssp z,\quad h \to O \ssp h,\quad \theta g h \to O \ssp \theta g h,\quad \theta g \to O \ssp \theta g \ssp O^\dagger, \quad O \in SU(4),
\end{equation}
which leaves $\beta$ and $\gamma$ invariant. It has an $U(2)$ subgroup which acts trivially on $z$ given in \eqref{cp3 vac g}. We use this symmetry to fix $\nu_2 = \nu_3 = 0$ and $\rho_3 = 0$. Thus, the dressed solution on $\bb{CP}^3$ is reduced to the one on $\bb{CP}^2$.

With the new parametrization, $u$ becomes
\begin{equation}
u = \( \frac{e^{-i \nu/2} \cos \rho_1}{\sqrt{2}} \,, \sin \rho_1 \,, 0,  \frac{e^{i \nu/2} \cos \rho_1}{\sqrt{2}} \)^{\rm T},
\label{cp3 u new}
\end{equation}
and $\beta,\ \gamma$ become
\begin{alignat}{3}
\beta &\equiv h^\dagger h & &= \sin^2 \rho_1 + \cos^2 \rho_1 \, \cos \[ \Sigma (\lambda_1) - \Sigma (\bar \lambda_1) \],
\label{cp3 beta} \\[1mm]
\gamma &\equiv h^\dagger \theta g h & &= \sin^2 \rho_1 - \cos^2 \rho_1 \, \cos \[ \Sigma (\lambda_1) + \Sigma (\bar \lambda_1) - \tau + \nu \],
\label{cp3 gamma}
\end{alignat}
where we take care of $\nu$-dependence for a later purpose.\footnote{See \eqref{condition unit radius} and below.} If we rewrite the spectral parameters as $\lambda_1 = e^{(i p + q)/2}$ and $\bar \lambda_1 = e^{(- i p + q)/2}$, the worldsheet coordinate $\Sigma (\lambda_1)$ becomes
\begin{equation}
\Sigma (\lambda_1) = \frac12 \( \tau - T \cos \alpha - i X \sin \alpha \),\qquad T = \frac{\tau - v \sigma}{\sqrt{1-v^2}} \,,\quad X = \frac{\sigma - v \tau}{\sqrt{1-v^2}} \,,
\label{def:Sigma}
\end{equation}
where
\begin{equation}
v = \frac{\lambda_1 + \bar \lambda_1}{\bar \lambda_1 \lambda_1 + 1} = \frac{\cos \( \frac{p}{2} \)}{\cosh \(\frac{q}{2}\)} ,\qquad
\tan \alpha = \frac{- i \(\lambda_1 - \bar \lambda_1\)}{\bar \lambda_1 \lambda_1 - 1} = \frac{\sin \( \frac{p}{2} \)}{\sinh \(\frac{q}{2}\)} \,.
\label{def:vw}
\end{equation}

\subsubsection{The solution}

\begin{figure}[htbp]
\begin{center}
\includegraphics[scale=0.8]{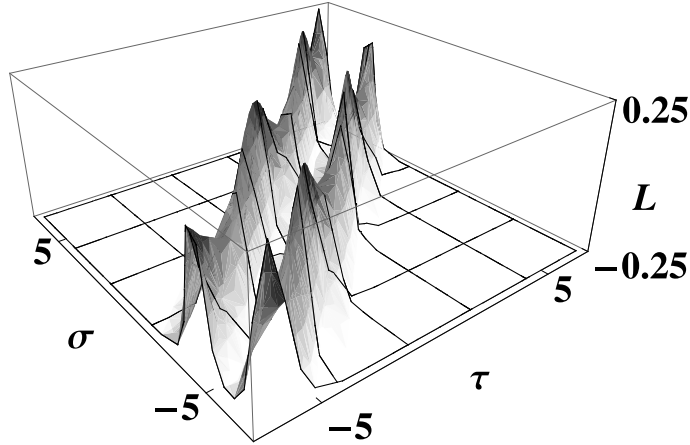}
\end{center}
\caption{Lagrangian density $L = \abs{D_\alpha z}^2$ for $\rho_1 = 0.9,\ \lambda_1 = \exp \( 0.6 i + 0.05 \)$. This figure describes a breather-like solution of $SU(3)/U(2)$ symmetric space sine-Gordon model.}
\label{fig:dzcp}
\end{figure}

The dressed solution is given by
\begin{alignat}{3}
\tilde z_1 &= \frac{e^{i \tau/2}}{2 \sqrt{2} \, \Lambda_z} \Biggl[ \cos^2 \rho_1 \left(- \frac{e^{X \sin \alpha} \lambda_1 + e^{-X \sin \alpha} \bar \lambda_1}{\lambda_1-\bar \lambda_1} + \frac{e^{-i T_\nu \cos \alpha} \left( e^{2 i T_\nu \cos \alpha} + \lambda_1 \bar \lambda_1 \right)}{\lambda_1 \bar \lambda_1-1}\right) \notag \\[1mm]
&\hspace{10cm} - \frac{2 \sin^2\rho_1 \left(\lambda_1^2-1\right) \bar \lambda_1}{(\lambda_1-\bar \lambda_1) (\lambda_1 \bar \lambda_1-1)} \Biggr] ,
\label{cp3 tz1} \\[2mm]
\tilde z_2 &= - \frac{\sin 2 \rho_1}{2 \Lambda_z} \, \cosh \( \frac{X \sin \alpha - i T_\nu \cos \alpha}{2} \),
\label{cp3 tz2} \\[1mm]
\tilde z_3 &= 0,
\label{cp3 tz3} \\
\tilde z_4 &= \frac{e^{-i \tau/2}}{2 \sqrt{2} \, \Lambda_z} \Biggl[ \cos^2 \rho_1 \left( - \frac{e^{-X \sin \alpha} \lambda_1 + e^{X \sin \alpha} \bar \lambda_1}{\lambda_1-\bar \lambda_1} + \frac{e^{-i T_\nu \cos\alpha} \left( 1 + e^{2 i T_\nu \cos \alpha} \lambda_1 \bar \lambda_1 \right)}{\lambda_1 \bar \lambda_1-1}\right) \notag \\[1mm]
&\hspace{10cm} -\frac{2 \sin^2\rho_1 \left(\lambda_1^2-1\right) \bar \lambda_1}{(\lambda_1-\bar \lambda_1) (\lambda_1 \bar \lambda_1-1)} \Biggr],
\label{cp3 tz4} \\[2mm]
\Lambda_z &= \Biggl\{ \lambda_1 \bar \lambda_1 \Biggl[ \( \frac{\sin^2 \rho_1 - \cos^2 \rho_1 \cos (T_\nu \cos \alpha)}{\bar \lambda_1 \lambda_1 - 1} \)^2 \notag \\[1mm]
&\hspace{65mm} - \( \frac{\sin^2 \rho_1 + \cos^2 \rho_1 \cosh (X \sin \alpha)}{\lambda_1-\bar \lambda_1} \)^2 \Biggr] \Biggr\}^{1/2},
\label{cp3 Lz}
\end{alignat}
where $T_\nu \cos \alpha \equiv T \cos \alpha - \nu$. They satisfy the boundary conditions
\begin{equation}
\tilde z_1 \; \to \; \frac{i \, e^{i \tau/2}}{\sqrt 2} \( \frac{\lambda_1}{\bar \lambda_1} \)^{\pm 1/2} ,\quad \tilde z_2, \, \tilde z_3 \; \to \; 0 ,\quad \tilde z_4 \; \to \; \frac{i \, e^{-i \tau/2}}{\sqrt 2} \( \frac{\bar \lambda_1}{\lambda_1} \)^{\pm 1/2} ,\quad \( \sigma \to \pm \infty \).
\end{equation}
in complete agreement with \eqref{CP3 boundary condition}. There is a symmetry
\begin{equation}
\tilde z_1 \( \frac{1}{\lambda_1} \,, \frac{1}{\bar \lambda_1} \) = - \tilde z_4 \( \lambda_1 \,, \bar \lambda_1 \) \quad {\rm at}\ \ \tau=0, \qquad \tilde z_2 \( \frac{1}{\lambda_1} \,, \frac{1}{\bar \lambda_1} \) = \tilde z_2 \( \lambda_1 \,, \bar \lambda_1 \),
\label{cp3 inversion symmetry}
\end{equation}
which looks similar to the inversion symmetry of quasi-momenta discussed in \cite{GV08a}. The gauge fields and the Lagrangian density are given by
\begin{alignat}{3}
A_\tau &= \frac{\cos^2 \rho_1 \cos \left(\frac{p}{2}\right) \cosh \left(\frac{q}{2}\right)}{2 \ssp \Lambda_z^2 \(\cosh q - \cos p\)} \( \frac{C_{X,+} \, \sin(T_\nu \cos \alpha)}{\sin \left(\frac{p}{2}\right) \cosh \left(\frac{q}{2}\right)} - \frac{C_{T,-} \, \sinh (X \sin \alpha)}{\cos \left(\frac{p}{2}\right) \sinh \left(\frac{q}{2}\right)} \),
\label{cp3 at} \\[2mm]
A_\sigma &= \frac{\cos^2 \rho_1 \cos \left(\frac{p}{2}\right) \cosh \left(\frac{q}{2}\right)}{2 \ssp \Lambda_z^2 \(\cosh q - \cos p\)} \( \frac{C_{T,-} \, \sinh (X \sin \alpha)}{\sinh \left(\frac{q}{2}\right) \cosh \left(\frac{q}{2}\right)} - \frac{C_{X,+} \, \sin (T_\nu \cos \alpha)}{\sin \left(\frac{p}{2}\right) \cos \left(\frac{p}{2}\right)} \), \label{cp3 as} \\[2mm]
\abs{D_a z}^2 &= -\frac{1}{32 \ssp \Lambda_z^4 \(\cosh q - \cos p\)} \, \times \notag \\[1mm]
&\Biggl[ \( \frac{C_{X,+}}{\sin \(\frac{p}{2}\)} \)^2 \left( 2 \sin^4\rho_1 - 6 \cos ^4 \rho_1 \cos^2 (T_\nu \cos \alpha) + C_{X,-}^2 + \frac{C_{X,+}^2 \sinh^2 \left(\frac{q}{2}\right)}{\sin^2 \(\frac{p}{2}\)} \right) \notag \\[1mm]
& + \(\frac{C_{T,-}}{\sinh \(\frac{q}{2}\)} \)^2 \( 2 \sin ^4 \rho_1 - 6 \cos^4 \rho_1 \cosh^2 (X \sin \alpha) + C_{T,+}^2 + \frac{C_{T,-}^2 \, \sin^2\(\frac{p}{2}\)}{\sinh^2 \(\frac{q}{2}\)} \) \Biggr],
\label{cp3 dzdz}
\end{alignat}
where
\begin{equation}
C_{T, \pm} \equiv \cos (T_\nu \cos \alpha) \cos^2\rho_1 \pm \sin^2\rho_1 \,, \quad
C_{X, \pm} \equiv \cosh (X \sin \alpha) \cos^2\rho_1 \pm \sin^2\rho_1 \,.
\end{equation}
They satisfy Lorenz gauge condition $\partial^\alpha A_\alpha = 0$. As discussed in Appendix \ref{sec:CPN Pohlmeyer}, the expressions \eqref{cp3 at}-\eqref{cp3 dzdz} defines breather-like solutions of $SU(3)/U(2)$ symmetric space sine-Gordon model, as shown in Figure \ref{fig:dzcp}.

\vfill

\newpage
Since the gauge fields are odd under parity transformation $( X, T_\nu ) \to (-X, - T_\nu )$, they do not contribute to the conserved charge \eqref{CP3 charges}. The conserved charges are evaluated as
\begin{equation}
E - \frac{J_1 - J_4}{2} = 4 h \sin \frac{p}{2} \cosh \frac{q}{2} = \frac{h}{i} \( \lambda_1 - \frac{1}{\lambda_1} - \bar \lambda_1 + \frac{1}{\bar \lambda_1} \),\quad J_1 + J_4 = J_2 = J_3 = 0,
\label{gen dispersion}
\end{equation}
which are independent of $\rho_1$\,. On the Chern-Simons side, these modes should correspond to excitations like
\begin{equation}
Y^1 (Y_4)^\dagger \quad \to \quad  Y^1 (Y_1)^\dagger,\ Y^2 (Y_2)^\dagger,\ Y^3 (Y_3)^\dagger,\ Y^4 (Y_4)^\dagger.
\end{equation}

\bigskip
Let us rewrite the dispersion relation \eqref{gen dispersion} as
\begin{equation}
E - \frac{J_1 - J_4}{2} = \sqrt{ n^2 + 16 h^2 \sin^2 \frac{p}{2}} \,,\qquad n \equiv 4 h \sin \frac{p}{2} \sinh \frac{q}{2} \,.
\label{gen dispersion 2}
\end{equation}
By repeating the argument we did in \eqref{Bohr-Sommerfeld condition}, one can show that $n$ is an action variable semiclassically quantized to integer. This situation is same as that of breather-like giant magnon solutions discussed in \cite{HM06}, so let us recall what we have learned in the \AdSxS\ case. In the paper \cite{HM06}, they rewrote the energy of a breather-like solution as the energy of a pair of elementary solutions, as
\begin{equation}
2 h' \sin \frac{p_1}{2} + 2 h' \sin \frac{p_2}{2} = \sqrt{{n'}^2 + 16 {h'}^2 \sin^2 \frac{p}{2}} \,,\quad n' \equiv 4 {h'} \sin \frac{p}{2} \sinh \frac{q}{2} \,,\quad {h'} \equiv \frac{\sqrt \lambda}{2 \pi} \,,
\end{equation}
with $p_1 = p + i q,\ p_2 = p - i q$, and $n'$ an action variable. This relation suggests that the semiclassical $S$-matrix should have single poles (or zeroes) at
\begin{equation}
\cos \( \frac{p_1}{2} \) - \cos \( \frac{p_2}{2} \) = - \frac{i n'}{4 h'} \quad \( n' \in \bb{Z}_{\ssp \ge 1} \).
\end{equation}
However, the full quantum theory of \AdSxS\ string predicts the existence of double poles at
\begin{equation}
u (p_1) - u (p_2) = - \frac{i n'}{2 h'} \quad \( n' \in \bb{Z}_{\ssp > 1} \), \qquad u (p) \equiv \frac{1}{2h'} \cot \frac{p}{2} \sqrt{1 + 4 {h'}^2 \sin^2 \frac{p}{2}} \,,
\end{equation}
where the branch of square root is carefully chosen \cite{BHL06, MS06, DHM07}. From this lesson, we may think of \eqref{gen dispersion 2} as giving the mean density of poles or zeroes (but not their exact location) for the $S$-matrix of \AdSxP\ string theory.\footnote{The author acknowledges to the referee of JHEP a cautious remark on this point.}

\subsubsection{Taking limits}

The dressed solution contains three real parameters ${\rm Re} \, \lambda_1 \,, {\rm Im} \, \lambda_1$ and $\rho_1$\,. The dressing matrix always becomes trivial in the limit ${\rm Im} \, \lambda_1 \to 0$, or $\rho_1 \to \pi/2$. So we concentrate on two other interesting limits, $\abs{\lambda_1} \to 1$ and $\rho_1 \to 0$.

As can be seen from \eqref{dress Q1} and \eqref{dress Q2}, the limit $\abs{\lambda_1} \to 1$, or equivalently $q \to 0$, forces the dressing matrix to be trivial unless $\gamma = 0$. Hence we send $\abs{\lambda_1}$ to $1$ under the constraint $\gamma = 0$, as explained below. First, we see from \eqref{cp3 gamma} that if $q=0$, the condition $\gamma=0$ is equivalent to
\begin{equation}
\gamma \equiv \sin^2 \rho_1 - \cos^2 \rho_1 \, \cos \( T_\nu \cos \alpha \) = \sin^2 \rho_1 - \cos^2 \rho_1 \, \cos \nu = 0.
\label{condition unit radius}
\end{equation}
We assume $\abs{\rho_1} \le \pi/4$ so that this equation has solutions with real $\nu$. Second, we expand the numerator of possibly divergent terms around $q=0$. This can be done by using
\begin{alignat}{3}
\nu &= \arccos \( \tan^2 \rho_1 \) + q \ssp \delta \nu + \cO (q^2),
\\[1mm]
e^{-i T_\nu \cos \alpha} &= - \frac{\(\sqrt{\cos (2 \rho_1)}-i\)^2}{2 \cos^2 (\rho_1)} \pare{ 1 - i q \left(\frac{T}{2 \sin \frac{p}{2}} - \delta \nu \right) + \cO (q^2) }.
\label{Tnu expand}
\end{alignat}
We can set $\delta \nu = 0$ by an appropriate shift of $T$.

There is no difficulty in taking the limit $\rho_1 \to 0$, and the $\bb{CP}^2$ solution reduces to that of $\bb{CP}^1$.

\subsubsection*{(1) $\bb{CP}^1$ limit}

When $\rho_1=0$, the dressed solution becomes
\begin{alignat}{3}
\tilde z_1 &= \frac{e^{i \tau/2}}{2 \sqrt{2} \, \Lambda_z} \[ -\frac{e^{X \sin \alpha} \lambda_1 + e^{-X \sin \alpha} \bar \lambda_1}{\lambda_1-\bar \lambda_1}+\frac{e^{-i T_\nu \cos \alpha} \left( e^{2 i T_\nu \cos \alpha} + \lambda_1 \bar \lambda_1 \right)}{\lambda_1 \bar \lambda_1-1} \], 
\label{cp1 tz1} \\[2mm]
\tilde z_4 &= \frac{e^{-i \tau/2}}{2 \sqrt{2} \, \Lambda_z} \[ - \frac{e^{-X \sin \alpha} \lambda_1 + e^{X \sin \alpha} \bar \lambda_1}{\lambda_1-\bar \lambda_1}+\frac{e^{-i T_\nu \cos\alpha} \left( 1 + e^{2 i T_\nu \cos \alpha} \lambda_1 \bar \lambda_1 \right)}{\lambda_1 \bar \lambda_1-1} \],
\label{cp1 tz4} \\[2mm]
\Lambda_z &= \Biggl\{ \lambda_1 \bar \lambda_1 \Biggl[ 
\( \frac{\cos (T_\nu \cos \alpha)}{\bar \lambda_1 \lambda_1 - 1} \)^2  - \( \frac{\cosh (X \sin \alpha)}{\lambda_1-\bar \lambda_1} \)^2 \Biggr] \Biggr\}^{1/2}.
\label{cp1 lambdaz}
\end{alignat}
with $\tilde z_2 = \tilde z_3 = 0$. They satisfy $z_1 = (z_4)^*$.

If we drop all $T_\nu$\ssp-dependent terms in the solution \eqref{cp1 tz1}-\eqref{cp1 tz4} by hand (rather than taking a limit) and set $\abs{\lambda_1}=1$, we obtain
\begin{equation}
\frac{\tilde z_1}{\tilde z_4} = e^{i \tau} \, \frac{\cosh \(X + \frac{i p}{2}\)}{\cosh \(X-\frac{ip}{2}\)} = e^{i \tilde \tau} \, \frac{\sinh \(\frac{\tilde X - i \tilde \omega}{2}\)}{\cosh \(\frac{\tilde X + i \tilde \omega}{2}\)} \, \quad \tilde \tau = \tau + \frac{\pi}{2} \,, \ \tilde X = \frac{X}{2} + \frac{i \pi}{4} \,,\ \tilde \omega = \frac{\pi}{2} - p \,,
\label{drop cp1gm}
\end{equation}
which coincides with the profile of $\bb{CP}^1$ giant magnons given in \eqref{cp1gm tz} with $\omega = \tilde \omega$.\footnote{The relation between $\omega$ and $p$ is actually incorrect, so \eqref{drop cp1gm} is not a consistent solution. This is because the coordinate $\tilde X$ of \eqref{drop cp1gm} is boosted by $v = \cos \frac{p}{2}$, while $X$ in \eqref{cp1gm tz} is by $v = \sin \omega$.} Although both solutions, \eqref{cp1 tz1}-\eqref{cp1 lambdaz} and \eqref{cp1gm tz}, have the pole at the same position $\lambda= \lambda_1 \,, \bar \lambda_1$ on the unit circle, the former carries twice as large charges as the latter.


\subsubsection*{(2) The limit $\bmt{\abs{\lambda_1} \to 1}$}

We carefully take the limit $\abs{\lambda_1} \to 1$ for general values of $\rho_1$ as described above. The string profile becomes
\begin{alignat}{3}
\tilde z_1 &= \frac{e^{i \tau/2}}{2 \sqrt{2} \, \Lambda_z} \Biggl[ - \cos^2 \rho_1 \( \frac{e^{X} \lambda_1 + e^{-X} \bar \lambda_1}{\lambda_1-\bar \lambda_1} \)
\notag \\[1mm]
&\hspace{5cm} + i \sqrt{\cos (2 \rho_1)} \( 1 + \frac{2 T}{\lambda_1 - \bar \lambda_1} \) - \sin^2 \rho_1 \( \frac{\lambda_1 + \bar \lambda_1}{\lambda_1 - \bar \lambda_1} \) \Biggr],
\label{unit tz1} \\[2mm]
\tilde z_2 &= - \frac{\sin \rho_1}{\sqrt{2} \Lambda_z} \[ \cosh \(\frac{X}{2}\) + i \sqrt{\cos (2 \rho_1)} \, \sinh \(\frac{X}{2}\) \],
\label{unit tz2} \\[1mm]
\tilde z_3 &= 0,
\label{unit tz3} \\
\tilde z_4 &= \frac{e^{-i \tau/2}}{2 \sqrt{2} \, \Lambda_z} \Biggl[ - \cos^2 \rho_1 \( \frac{e^{-X} \lambda_1 + e^{X} \bar \lambda_1}{\lambda_1-\bar \lambda_1} \)
\notag \\[1mm]
&\hspace{5cm} - i \sqrt{\cos (2 \rho_1)} \( 1 - \frac{2 T}{\lambda_1 - \bar \lambda_1} \) - \sin^2 \rho_1 \( \frac{\lambda_1 + \bar \lambda_1}{\lambda_1 - \bar \lambda_1} \) \Biggr],
\label{unit tz4} \\[2mm]
\Lambda_z &= \pare{ \frac{-1}{( \lambda_1-\bar \lambda_1 )^2} \[ \( \cos^2 \rho_1 \, \cosh X + \sin^2 \rho_1 \)^2 + \cos 2 \rho_1 \, T^2 \] }^{1/2}.
\label{unit Lz}
\end{alignat}
The gauge fields and the Lagrangian density are given by
\begin{alignat}{3}
A_\tau &= - \frac{\sqrt{\cos (2 \rho_1)}}{4 \sin^3 \(\frac{p}{2}\) \Lambda_z^2} \, \Biggl[ T \sinh (X) \cos ^2(\rho_1)+\cos \left(\frac{p}{2}\right) \left(\cosh (X) \cos ^2(\rho_1)+\sin^2(\rho_1)\right)  \Biggr],
\label{unit at} \\[2mm]
A_\sigma &= \frac{\sqrt{\cos (2 \rho_1)}}{4 \sin^3 \(\frac{p}{2}\) \Lambda_z^2} \, \Biggl[ T \sinh (X) \cos ^2(\rho_1) \cos \left(\frac{p}{2}\right) + \cosh (X) \cos ^2(\rho_1)+\sin^2(\rho_1) \Biggr],
\label{unit as} \\[1mm]
\abs{D_a z}^2 &= \frac{-1}{64 \sin^4 \(\frac{p}{2}\) \Lambda_z^4} \, \Biggl[ T^4 \cos ^2(2 \rho_1) - 6 \ssp T^2 \cos (2 \rho_1) \left(\cos^4(\rho_1) \cosh ^2(X)-\sin ^4(\rho_1)\right)
\notag \\
&\hspace{23mm} + \left(\cos^2(\rho_1) \cosh (X)-3 \sin ^2(\rho_1)\right) \left(\cosh (X) \cos ^2(\rho_1)+\sin ^2(\rho_1)\right)^3 \Biggr].
\label{unit dzdz}
\end{alignat}
They satisfy the Lorenz gauge condition, $\partial^\alpha A_\alpha = 0$. Note that the gauge fields are even under parity transformation.

Since the gauge fields are proportional to $\sqrt{\cos (2 \rho_1)}$, the parameter $\rho_1$ must take values in between $[-\pi/4, \pi/4]$, as is expected from the condition \eqref{condition unit radius}. When $\rho_1 = \pm \pi/4$, they become equivalent to the $\bb{RP}^2$ giant magnon solutions \eqref{DGM on RP3}, after rescaling of $(\tau, \sigma)$. In terms of the reduced sine-Gordon system, it implies that the breather of a sine-Gordon model ($\rho_1=0$) is continuously connected to the kink of another sine-Gordon model ($\rho_1=\pi/4,\ \abs{\lambda_1} = 1$).

\subsection{Comments on other embeddings}

As noticed in \cite{SV06, KSV06}, the dressings of different coset spaces give us different soliton solutions. For the present case, the $\bb{RP}^3$ dyonic giant magnons can be easily obtained by dressing $SU(2) \times \overline{SU(2)}$. The dressing of the BPS vacuum on $SU(4)/U(3)$ with rank one projector cannot reproduce such solutions.\footnote{The term ``rank" refers to the rank of constant vector $u$. See footnote \ref{note:rank}.}

We may consider dressing the $SO(6)/U(3)$ model instead of $SU(4)/U(3)$. As discussed in Appendix \ref{sec:so(6)/u(3)}, however, it turns out that the dressing matrix becomes trivial when the spectral parameters approach the unit circle. We are unable to construct ``dyonic giant magnons" on $\bb{CP}^3$ neither in this way.

\section{Summary and Discussion}\label{sec:summary}

In this paper, we consider the classical string spectrum of \RCP{3}\ sigma model in the decompactification limit. We constructed a family of giant magnon solutions with $SU(2)$ symmetry by means of the dressing method on $SU(4)/U(3)$. All such solutions obey the same square-root type dispersion relation which is, at least na\"ively, expected from the BPS relation of the centrally extended $psu(2|2)$ symmetry.

The new solutions are neutral with respect to the global charges of $psu(2|2)$, and thus they could be non-BPS boundstates which receive quantum corrections. It is known that there are no non-BPS boundstates in the \AdSxS\ case, in the sense that neutral states are equivalent to a composite of two oppositely-charged dyonic giant magnons \cite{SV06}. Since different boundstate spectrum should lead to different singularity structure of the worldsheet $S$-matrix \cite{DHM07, DO07}, it is interesting to determine whether our solutions are BPS or not.

There remains a problem to construct an explicit profile of dyonic giant magnon solutions. We expect to find ways to construct such soliton solutions with the help of classical integrability. One direction is to study the reduced sine-Gordon system discussed in Appendix \ref{sec:CPN Pohlmeyer}. Another direction is to study in detail classical membrane spectrum in AdS${}_4 \times {}$S${}^7/\bb{Z}_k$ for general $k$, and carefully take the limit $k \to \infty$ limit \cite{Bozhilov02, Bozhilov07a, BR07, AB08d}. We hope to revisit such problems in future.

\subsubsection*{Acknowledgments}

I would like to acknowledge Dmitri Bykov, Sergey Frolov and Yasuyuki Hatsuda for discussions. This work is supported by the Science Foundation Ireland under Grant No.07/RFP/PHYF104.

\appendix

\section{Pohlmeyer reduction on $\bmt{\bb{CP}^N}$}\label{sec:CPN Pohlmeyer}

We revisit the reduction problem of classical strings on \RCP{N}. It is known that equations of motion on $\bb{CP}^N \simeq SU(N+1)/U(N)$ can equivalently be rewritten as symmetric space sine-Gordon equations \cite{DRS80a, DRS80b, EH81, BPS95, FGHM96, Miramontes08}. We would like to clarify an explicit relation between the $\bb{CP}^N$ coordinates and the sine-Gordon fields in order to relate the solutions of two theories. They will also help us to construct new solutions of the $\bb{CP}^N$ sigma model as in \cite{CDO06a, OS06}.

The reduction procedure goes parallel with the \RS{N}\ case done by Pohlmeyer, which can be outlined as follows. One chooses a $U(1)$-invariant, orthonormal basis of the tangent space of $\bb{C}^{N+1}$, say $\{ \vec v^{\, 1}, \cdots , \vec v^{\, N+1} \}$. Then one differentiate the basis vectors. The result can be expanded by the basis itself as $\partial_\alpha \vec v^{\, k} = M_\alpha \cdot \vec v^{\, k}$. The compatibility condition $\(D_\alpha D_\beta - D_\beta D_\alpha \) \vec v^{\, k} = - i F_{\alpha \beta} \, \vec v^{\, k}$ gives differential equations for the matrix elements of $M_\alpha$. One can recast them into sine-Gordon like equations through appropriate parametrization of $M_\alpha$.

\subsection{Constraints and Identities}\label{sec:constraints}

Let us define light-cone coordinates by $\partial_{\pm} = \partial_{\tau} \pm \partial_{\sigma}$. The energy-momentum conservation becomes $\partial_+ T_{--} = \partial_- T_{++} = 0$, and we can rewrite Virasoro constraints as
\begin{equation}
\frac14 = \abs{D_+ z_i}^2 = \abs{D_- z_i}^2 ,
\label{P Virasoro}
\end{equation}
where we rescaled worldsheet coordinates to set $\kappa = 1$. We introduce a dynamical degree of freedom $u$ and rewrite the equations of motion as
\begin{alignat}{3}
\frac{\cos u}{2} &= \overline{D_+ z_i} \, D_- z_i + D_+ z_i \, \overline{D_- z_i} \,,
\label{P reduction} \\[2mm]
- \frac{\cos u}{2} \, z_i &= D_+ D_- z_i + D_- D_+ z_i \,.
\label{P eom}
\end{alignat}
We can derive some identities from $\bar z_i \, z_i = 1$ and \eqref{U1 connection},
\begin{alignat}{3}
0 &= \bar z_i \, D_\alpha z_i \,,
\label{P zDz} \\
F_{+-} &= - i \( \overline{D_+ z_i} \, D_- z_i - D_+ z_i \, \overline{D_- z_i} \).
\label{P Fpm}
\end{alignat}
By combining \eqref{P Fpm} with \eqref{P reduction}, we find
\begin{equation}
\overline{D_+ z_i} \, D_- z_i = \frac{\cos u + 2 i F_{+-}}{4} \,, \qquad \overline{D_- z_i} \, D_+ z_i = \frac{\cos u - 2 i F_{+-}}{4} \,,
\label{P zpzm}
\end{equation}
and \eqref{P Fpm} with \eqref{P eom},
\begin{equation}
D_+ D_- z_i = - \( \frac{\cos u + 2 i F_{+-}}{4} \) z_i \,. 
\label{P DpDmz}
\end{equation}

\bigskip
We differentiate the above equations to find identity relations. Derivative of \eqref{P zDz} gives
\begin{equation}
- \frac14 = \bar z_i \, D_+^2 z_i = \bar z_i \, D_-^2 z_i \,.
\label{P constraints 1}
\end{equation}
Derivative of Virasoro constraints \eqref{P Virasoro} gives
\begin{equation}
0 = \overline{D_\pm^2 z_i} \, D_\pm z_i + D_\pm^2 z_i \, \overline{D_\pm z_i} \,.
\label{P constraints 2}
\end{equation}
From \eqref{P zpzm}, we obtain
\begin{equation}
\overline{D_+ z_i} \, D_-^2 z_i = \partial_- \( \frac{\cos u + 2 i F_{+-}}{4} \), \qquad
\overline{D_- z_i} \, D_+^2 z_i = \partial_+ \( \frac{\cos u - 2 i F_{+-}}{4} \),
\label{P constraints pmm}
\end{equation}
and
\begin{equation}
4 \ssp \overline{D_+^2 z_i} \, D_-^2 z_i = \partial_+ \partial_- \( \cos u + 2 i F_{+-} \) + \frac14 \( \cos u + 2 i F_{+-} \) \(\cos u + 6 i F_{+-} \).
\label{P constraints ppmm}
\end{equation}

\bigskip
We introduce another dynamical degrees of freedom $H_\pm$ by
\begin{equation}
H_\pm \equiv - i \( \overline{D_\pm^2 z_i} \, D_\pm z_i - D_\pm^2 z_i \, \overline{D_\pm z_i} \),
\label{def:Hpm}
\end{equation}
so that we have
\begin{equation}
\overline{D_\pm z_i} \, D_\pm^2 z_i = - \frac{i}{2} \, H_\pm \,,
\label{P constraints mpp}
\end{equation}
where we used \eqref{P constraints 2}. It follows that
\begin{equation}
\partial_+ H_- = - \partial_- H_+ = F_{+-} \,.
\label{P constraints HF}
\end{equation}
Thus, $H_\pm/2$ coincides with the $U(1)$ gauge field $A_\pm$ satisfying Lorenz gauge condition $\partial^\alpha A_\alpha = 0$. Let us introduce a new variable $\varphi$ by
\begin{equation}
H_\pm \equiv \mp \partial_\pm \varphi, \qquad \partial_+ \partial_- \varphi = F_{+-} \,.
\label{def:varphi}
\end{equation}

\subsection{Reduction procedure}

By using $\bb{CP}^N \subset \bb{C}^{N+1}$, we expand the second-order covariant derivatives as\footnote{We omit the subscript $i$ from $z_i$ below.}
\begin{align}
D_+^2 z &= a_1 \, z + a_2 \, D_+ z + a_3 \, D_- z + \sum_{j=4}^{N+1} a_j \, v^j \,, \notag \\
D_-^2 z &= b_1 \, z + b_2 \, D_- z + b_3 \, D_+ z + \sum_{j=4}^{N+1} b_j \, v^j \,,
\label{DDz expand CPN}
\end{align}
where $v^{\, j}$ are gauge-invariant basis vectors, satisfying orthonormal conditions
\begin{equation}
0 = \bar v^j z = \bar v^j D_\pm z \,, \qquad \delta^{jk} = \bar v^j \, v^k .
\label{orthonormal}
\end{equation}

The coefficients $a_1$ and $b_1$ are determined by the equations \eqref{P constraints 1} as
\begin{equation}
a_1 = b_1 = - \frac14 \,.
\label{coeff a1b1}
\end{equation}
The coefficients $a_2, a_3$ are constrained by the equations \eqref{P constraints pmm} and \eqref{P constraints mpp} as
\begin{equation}
\begin{pmatrix}
\cos u - 2 i F_{+-} & 1 \\
1 & \cos u + 2 i F_{+-} 
\end{pmatrix} \( \begin{array}{*{20}c}
a_2 \\
a_3 
\end{array} \) = \( \begin{array}{*{20}c}
\partial_+ \( \cos u - 2 i F_{+-} \) \\
2 \ssp i \ssp \partial_+ \varphi
\end{array} \).
\label{coeffs a2a3}
\end{equation}
Constraints for $b_2, b_3$ are given by
\begin{equation}
\begin{pmatrix}
\cos u + 2 i F_{+-} & 1 \\
1 & \cos u - 2 i F_{+-} 
\end{pmatrix} \( \begin{array}{*{20}c}
b_2 \\
b_3 
\end{array} \) = \( \begin{array}{*{20}c}
\partial_- \( \cos u + 2 i F_{+-} \) \\
- 2 \ssp i \ssp \partial_- \varphi
\end{array} \).
\label{coeffs b2b3}
\end{equation}

\bigskip
If the $2 \times 2 $ matrix in \eqref{coeffs a2a3} is not degenerate, that is $F_{+-} \neq \pm \sin (u)/2$, then this equation is solved by
\begin{equation}
\( \begin{array}{*{20}c}
a_2 \\
a_3 
\end{array} \) = \frac{1}{\bar \eta \eta - 1} \begin{pmatrix}
\eta & - 1 \\
- 1 & \bar \eta
\end{pmatrix}  \( \begin{array}{*{20}c}
\partial_+ \bar \eta \\
2 \ssp i \ssp \partial_+ \varphi
\end{array} \),
\label{solution a2a3}
\end{equation}
where $\eta = \cos u + 2 i F_{+-}$ and $\bar \eta = \cos u - 2 i F_{+-}$\,. We can evaluate the left hand side of \eqref{P constraints ppmm} as
\begin{alignat}{3}
4 \ssp \overline{D_+^2 z_i} \, D_-^2 z_i &= 4 \ssp \bar a_1 b_1 + \bar a_2 b_3 + \bar a_3 b_2 + \eta \, \bar a_2 b_2 + \bar \eta \, \bar a_3 b_3 + 4 \sum_{j=4}^{N+1} \bar a_j b_j \,,
\label{reduced eq 1} \\
&= \frac14 + \frac{\bar \eta \, \partial_+ \eta \, \partial_- \eta - 4 \eta \, \partial_+ \varphi \, \partial_- \varphi + 2 i \( \partial_- \eta \, \partial_+ \varphi + \partial_+ \eta \, \partial_- \varphi \)}{\bar \eta \eta - 1} + 4 \sum_{j=4}^{N+1} \bar a_j b_j \,.
\label{reduced eq 2}
\end{alignat}

\bigskip
Suppose now that the $2 \times 2 $ matrix in \eqref{coeffs a2a3} is degenerate. We may set $F_{+-} = \sin (u)/2$, by the flip of $u \mapsto -u$ if necessary. The equation \eqref{coeffs a2a3} tells us
\begin{alignat}{3}
a_3 &= - \( a_2 + i \ssp \partial_+ u \) e^{-i u}, &\qquad \partial_+ \varphi &= - \frac{\partial_+ u}{2} \,,
\label{deg:a3} \\[1mm]
b_3 &= - \( b_2 - i \ssp \partial_- u \) e^{i u}, &\qquad \partial_- \varphi &= - \frac{\partial_- u}{2} \,.
\label{deg:b3}
\end{alignat}
The condition \eqref{P constraints HF} gives
\begin{equation}
0 = - \partial_+ \partial_- u - \, \sin u = \partial_a^2 u - \sin u \,,
\label{sine-Gordon eq}
\end{equation}
which is sine-Gordon equation.

For our reduction procedure to be consistent, the relation \eqref{P constraints ppmm} must reduce to the same sine-Gordon equation as \eqref{sine-Gordon eq}. This can be checked by evaluating \eqref{reduced eq 1} by using \eqref{deg:a3} and \eqref{deg:b3}. The result is
\begin{equation}
4 \ssp \overline{D_+^2 z_i} \, D_-^2 z_i = \frac14 - e^{iu} \, \partial_+ u \, \partial_- u + 4 \sum_{j=4}^{N+1} \bar a_j b_j \,,
\label{reduced eq 3}
\end{equation}
which is independent of undetermined coefficients $a_2$ and $b_2$\,. We will see later that the coefficients $a_j, b_j\ (j \ge 4)$ vanish in this case. Thus the second-order differential equation \eqref{P constraints ppmm} is equivalent to \eqref{sine-Gordon eq}.

\subsection{$\bb{CP}^1$ case}

Recall that $\bb{CP}^1$ is locally isomorphic to ${\rm S}^2$, and the Pohlmeyer reduction of \RS{2}\ sigma model provides us the sine-Gordon equation. From this reasoning we can fix the normalization of sine-Gordon coupling. We find
\begin{equation}
- \cos u = 4 \abs{D_\alpha z_i}^2,
\label{CP1 Pohlmeyer 1}
\end{equation}
where $u (\tau, \sigma)$ satisfies the sine-Gordon equation $\partial_\alpha^2 u - \sin u = 0$.

This equation has the same normalization as the one obtained for the degenerate case \eqref{sine-Gordon eq}. This result is expected. Since the tangent space of $\bb{C}^2$ is two-dimensional, the set of equations \eqref{coeffs a2a3} must be overdetermined and the expansion \eqref{DDz expand CPN} must contain a redundant parameter. In fact, the degeneracy condition $F_{+-} = \pm \sin (u)/2$ is identically satisfied on the $\bb{CP}^1$ space.

\subsection{Nondegenerate $\bb{CP}^2$ case}\label{sec:CP2 sG}

Since the tangent space of $\bb{C}^3$ is three-dimensional, we can set $a_j = b_j = 0$ for $j \ge 4$ in \eqref{reduced eq 2}. The differential equation \eqref{P constraints ppmm} becomes
\begin{equation}
- \partial_+ \partial_- \eta - \frac{\bar \eta \, \partial_+ \eta \, \partial_- \eta - 4 \eta \, \partial_+ \varphi \, \partial_- \varphi + 2 i \( \partial_- \eta \, \partial_+ \varphi + \partial_+ \eta \, \partial_- \varphi \)}{1 - \bar \eta \eta} + \frac{1 - 2 \eta^2 + \bar \eta \eta}{4} = 0.
\label{sgcp2 eom 1}
\end{equation}
We redefine $\eta = \zeta \, e^{-2i\varphi}$ and $\bar \eta = \bar \zeta \, e^{2i\varphi}$, and rewrite this equation by using
\begin{equation}
- 4 i \partial_a^2 \varphi = \eta - \bar \eta = \zeta \, e^{-2i\varphi} - \bar \zeta \, e^{2i\varphi},
\label{eom varphi}
\end{equation}
as
\begin{equation}
\partial_a^2 \zeta + \frac{\bar \zeta (\partial_a \zeta )^2}{1 - \bar \zeta \zeta} + \( \frac{1 - \bar \zeta \zeta}{4} \) e^{2i\varphi} = 0.
\label{sgcp2 eom 2}
\end{equation}
All of the equations \eqref{eom varphi}, \eqref{sgcp2 eom 2}, and the complex conjugate of \eqref{sgcp2 eom 2}, can be derived from the Lagrangian
\begin{equation}
{\cal L} = \frac{\partial^a \bar \zeta \partial_a \zeta}{1 - \bar \zeta \zeta} - \frac{\bar \zeta \, e^{2 i \varphi} + \zeta \, e^{-2i\varphi}}{4} + (\partial_a \varphi)^2 .
\label{sgcp2 Lag}
\end{equation}

One can reproduce the sine-Gordon equation for $\bb{CP}^1$ by setting $\zeta = 1$. This solves the equation \eqref{sgcp2 eom 2}, and the constraint \eqref{eom varphi} gives
\begin{equation}
0 = \partial_a^2 \varphi - \frac12 \, \sin 2 \varphi,
\label{cp2 to cp1 sG}
\end{equation}
which is \eqref{sine-Gordon eq} with $u = - 2 \varphi$. If we set $\varphi = 0$, that is $\bar \zeta = \zeta = \cos u$, the equation \eqref{sgcp2 eom 2} becomes
\begin{equation}
\partial_a^2 \zeta + \frac{\zeta \( \partial_a \zeta \)^2}{1-\zeta^2} + \frac{1 - \zeta^2}{4} = 0,
\label{sine-Gordon eq 2}
\end{equation}
which is again sine-Gordon, with the different normalization from \eqref{cp2 to cp1 sG}.

If we define three real variables $\alpha, \beta, \gamma$ by
\begin{equation}
\eta = - \cos \alpha \, e^{i \gamma}, \qquad \varphi = \frac{\beta}{2} \,, \qquad \zeta = \eta \, e^{2i\varphi} = - \cos \alpha \, e^{i (\gamma + \beta)},
\label{def:abg}
\end{equation}
we can rewrite \eqref{sgcp2 Lag} as
\begin{equation}
{\cal L} = (\partial_a \alpha )^2 + \cot^2 \alpha \, \partial_a \( \gamma + \beta \) \partial^a \( \gamma + \beta \) + \frac12 \cos \alpha \cos \gamma + \frac14 \( \partial_a \beta \)^2,
\label{sgcp2 Lag 2}
\end{equation}
which is the Lagrangian obtained in \cite{EH81}. The explicit relation between sine-Gordon fields $\alpha, \beta, \gamma$ and $\bb{CP}^2$ coordinates has been mentioned in \cite{EH81}, which agree with ours \eqref{P zpzm}, \eqref{P constraints mpp} up to $\gamma \to \gamma + \pi$.

\subsection{Degenerate $\bb{CP}^N$ cases}

We will show that the coefficients $a_j, b_j\ (j \ge 4)$ always vanish in the degenerate case. Let us expand the second-order covariant derivatives as
\begin{align}
D_+^2 z_i &= a_1 \, z_i + a_2 \, D_+ z_i + a_3 \, D_- z_i + \sum_{j=4}^{N+1} a_j \, v_i^{(j)} \,, \notag \\
D_-^2 z_i &= b_1 \, z_i + b_2 \, D_- z_i + b_3 \, D_+ z_i + \sum_{j=4}^{N+1} b_j \, v_i^{(j)} \,,
\end{align}
where $v^{\; (j)}$ satisfy the orthonormal conditions \eqref{orthonormal}. We omit the indices $i$ below.

By taking derivatives of \eqref{orthonormal} and using identities in Section \ref{sec:constraints}, we obtain
\begin{equation}
\overline{D_\alpha v^{(j)}} \, z = 0, \qquad \bar v^j D_\alpha D_\beta z + \overline{D_\alpha v^j} \, D_\beta z = 0.
\label{der ON}
\end{equation}
We can rewrite the coefficients $a_j$ and $b_j$ as
\begin{equation}
a_j = \bar v^j D_+^2 z = - \overline{D_+ v^j} \, D_+ z ,\qquad b_j = \bar v^j D_-^2 z = - \overline{D_- v^j} \, D_- z .
\label{def:ajbj}
\end{equation}
We also find
\begin{equation}
\overline{D_\pm v^j} \, D_\mp z = - \bar v^j D_\pm D_\mp z = \( \frac{\cos u + 2 i F_{+-}}{4} \) \bar v^j z = 0.
\label{vpzm}
\end{equation}

Let us expand the covariant derivatives of the basis vectors as
\begin{align}
\overline{D_+ v} &= r_1 \, \bar z + r_2 \, \overline{D_+ z} + r_3 \, \overline{D_- z} + \sum_{j=4}^{N+1} r_j \, \bar v^j \,, \notag \\
\overline{D_- v} &= s_1 \, \bar z + s_2 \, \overline{D_- z} + s_3 \, \overline{D_+ z} + \sum_{j=4}^{N+1} s_j \, \bar v^j \,.
\label{Dv expand CPN}
\end{align}
The first equation of \eqref{der ON} shows $r_1 = s_1 = 0$. The relations \eqref{def:ajbj} and \eqref{vpzm} give us
\begin{equation}
\begin{pmatrix}
1 & \cos u + 2 i F_{+-} \\
\cos u - 2 i F_{+-} & 1 \\
\end{pmatrix} \( \begin{array}{*{20}c}
r_2 \\
r_3 
\end{array} \) = \( \begin{array}{*{20}c}
- a_j \\
0
\end{array} \).
\label{coeffs r2r3}
\end{equation}
and similar equations for $b_j$\,. When this $2 \times 2$ matrix is degenerate, that is when $F_{+-} = \pm \sin (u)/2$, we obtain $a_j = b_j = 0$ for $j \ge 4$.

\bigskip
For the non-degenerate case, one can derive differential equations for $a_j$ and $b_j$ from the compatibility condition $\( D_+ D_- - D_- D_+ \) v^j = - i F_{+-} v^j$. We do not discuss them here because they look complicated. We just refer to \cite{Rashkov08}, which studied the reduced sine-Gordon type equations of \AdSxP.

\section{On dressing $\bmt{SO(6)/U(3)}$}\label{sec:so(6)/u(3)}

We will show that the dressing matrix of the $SO(6)/U(3)$ model becomes trivial when the spectral parameters are on the unit circle following \cite{HSS84a, HSS84b}.

In the $SO(6)/U(3)$ model, the minimum set of poles in the dressing matrix is $\lambda_1 \,, 1/\lambda_1 \,, \bar \lambda_1 \,, 1/\bar \lambda_1$ when $\abs{\lambda_1} \neq 1$. We write the dressing matrix $\chi$ and $\chi^{-1}$ as
\begin{alignat}{3}
\chi (\lambda) &= {\bf 1}_6 + \frac{Q_1}{\lambda - \lambda_1} + \frac{Q_{\hat 1}}{\lambda - 1/\lambda_1} + \frac{Q_{\bar 1}}{\lambda - \bar \lambda_1} + \frac{Q_{\hat{\bar 1}}}{\lambda - 1/\bar \lambda_1} \,,\\[1mm]
\chi^{-1} (\lambda) &= {\bf 1}_6 + \frac{R_1}{\lambda - \lambda_1} + \frac{R_{\hat 1}}{\lambda - 1/\lambda_1} + \frac{R_{\bar 1}}{\lambda - \bar \lambda_1} + \frac{R_{\hat {\bar 1}}}{\lambda - 1/\bar \lambda_1} \,,
\end{alignat}
where $Q_i = X_i F_i^\dagger$ and $R_i = H_i K_i^\dagger$. Since $\chi$ and $\chi^{-1}$ share the same pole, we have to impose $F_i^\dagger H_i = 0$ and consider
\begin{equation}
\Gamma_{ii} = - F_i^\dagger \, \psi' (\lambda_i) \ssp \psi^{-1} (\lambda_i) H_i + f_i \ssp c_i \ssp h_i \,.
\end{equation}

There are three constraints imposed on $\psi (\lambda)$,
\begin{equation}
\[ \psi (\bar \lambda) \]^\dagger = \psi^{-1} (\lambda), \qquad
\[ \psi (\bar \lambda) \]^* = \psi (\lambda), \qquad
g K \, \psi (\lambda) \, K = \psi (1/\lambda),
\end{equation}
where $g \equiv \psi (0)$ and $K$ is an antisymmetric involution defined in \eqref{def:YK}. The first, unitarity constraint is solved by\footnote{$\lambda_{\bar \imath} = \bar \lambda_i$ and $\lambda_{\hat \imath} = 1/\lambda_i$\,.}
\begin{equation}
F_i = H_{\bar \imath} \,,\qquad \Gamma_{ii} = - \( \Gamma_{\bar \imath \bar \imath} \)^\dagger \,.
\label{unitary cns}
\end{equation}
The second, orthogonality constraint by
\begin{equation}
F_i = \( F_{\bar \imath} \)^* \,,\qquad  \Gamma_{ii} = \( \Gamma_{\bar \imath \bar \imath} \)^* \,.
\label{orthogonal cns}
\end{equation}
The third, inversion constraint by
\begin{equation}
F_i = g K \, F_{\hat \imath} \,,\qquad  \Gamma_{ii} = - \lambda_i^2 \, \Gamma_{\hat \imath \hat \imath} \,.
\label{inversion cns}
\end{equation}
Two conditions \eqref{unitary cns} and \eqref{orthogonal cns} shows
\begin{equation}
F_i^{\rm T} F_i = f_i^{\rm T} f_i = 0, \qquad \Gamma_{ii}^{\rm T} = - \Gamma_{ii} \,.
\label{self orthogonal}
\end{equation}
It follows that $\Gamma_{ii}=0$ when $\Gamma_{ii}$ is a rank-one matrix.

If $\abs{\lambda_1} = 1$, we have to solve $F_i^* = g K \, F_i$\,, which is equivalent to $f_i^* = K f_i$\,. Since $K$ is a real antisymmetric matrix, the conditions $f_i^* = K \, f_i$ means $0 = f_i^{\rm T} K f_i = - f_i^\dagger f_i$\,, namely $f_i = 0$. Therefore, the dressing matrix becomes trivial on the unit circle.\footnote{Considering the fact that the overall normalization of $f_i$ is unimportant in the dressing method, it might be possible to obtain a nontrivial profile in the limit $\abs{\lambda_1} \to 1$, if one can take such limit in a sophisticated way.}

\end{document}